# MDiff-FMT: Morphology-aware Diffusion Model for Fluorescence Molecular Tomography with Small-scale Datasets

Peng Zhang[1], Qianqian Xue[1], Xingyu Liu[1], Guanglei Zhang[2], Wenjian Wang[1*], Jiye Liang[1*]

[1]School of Computer and Information Technology, Shanxi University, Taiyuan 030006, China
[2]School of Biological and Medical Engineering, Beihang University, Beijing 100191, China
wjwang@sxu.edu.cn, ljy@sxu.edu.cn

**Abstract.** Fluorescence molecular tomography (FMT) is a sensitive optical imaging technology widely used in biomedical research. However, the ill-posedness of the inverse problem poses a huge challenge to FMT reconstruction. Although end-to-end deep learning algorithms have been widely used to address this critical issue, they still suffer from high data dependency and poor morphological restoration. In this paper, we report for the first time a morphology-aware diffusion model, MDiff-FMT, based on denoising diffusion probabilistic model (DDPM) to achieve high-fidelity morphological reconstruction for FMT. First, we use the noise addition of DDPM to simulate the process of the gradual degradation of morphological features, and achieve fine-grained reconstruction of morphological features through a stepwise probabilistic sampling mechanism, avoiding problems such as loss of structure details that may occur in end-to-end deep learning methods. Additionally, we introduce the conditional fluorescence image as structural prior information to sample a high-fidelity reconstructed image from the noisy images. Numerous numerical and real phantom experimental results show that the proposed MDiff-FMT achieves SOTA results in morphological reconstruction of FMT without relying on large-scale datasets. The code is available at ***.

## 1 Introduction

Fluorescence molecular tomography (FMT), as a sensitive optical molecular imaging technique, has been applied in a variety of medical fields such as drug development [1,2], early diagnosis of diseases[3,4] and treatment monitoring[5,6]. However, due to the strong scattering effect of biological tissues and insufficient surface measurement data, the reconstruction performance of FMT is greatly limited due to its severe ill-posedness. Therefore, it is urgent to explore effective reconstruction algorithms to achieve high-quality FMT reconstruction.

Previously, researchers have proposed various reconstruction algorithms, the most representative of which are iteration-based traditional regularization methods. These methods primarily alleviate the ill-posedness by introducing regularization terms such as L2 norm [7], L1 norm [8,9,10], L2,1 norm [11]. Although they can enhance the image quality to a certain extent, their performance is heavily relying on hyperparameter selection, and typically suffer from the problems long inference time and low reconstruction efficiency.

In recent years, data-driven deep learning methods [12,13,14,15,16] have been widely used in FMT reconstruction, which mainly establish the mapping relationship between fluorescence signals and reconstructed images through end-to-end network models. These algorithms can effectively solve the problems of long time and low reconstruction efficiency of traditional FMT reconstruction algorithms, but they require a large number (usually tens of thousands) of high-quality paired data for training. Afterwords, model-driven deep learning methods have also been gradually applied to FMT reconstruction, usually by unfolding the iterative algorithmic process into a deep network to achieve FMT reconstruction under the condition of domain prior knowledge [17,18,19,20,21]. These algorithms can improve model reconstruction performance using relatively few training samples, but the computational complexity is relatively large and not fine enough in morphological structure restoration.

The diffusion model [22] is an emerging generative model that has shown strong performance in medical imaging due to its unique forward diffusion-inverse denoising process. Unlike CNN's pixel-by-pixel feature extraction or Transformer's global attention, the diffusion model's noise phase can generate diverse samples and simulate various interference factors in medical image acquisition. Additionally, each step of its inverse stage estimates the data distribution probability to reconstruct images, enabling progressive learning of essential image features and enhancing model morphological mining capabilities. This allows the model to achieve superior results even with limited training samples. Inspired by the powerful performance of the diffusion model, we propose a morphology-aware diffusion model for FMT reconstruction, termed MDiff-FMT. The key contributions of this work are as follows:

• We propose, for the first time, a novel morphology-aware diffusion network model for 3D FMT reconstruction. Unlike conventional CNN or Transformer-based approaches, MDiff-FMT eliminates the need for large-scale training datasets and can progressively generate high-quality 3D reconstruction images from random noise, even with limited training data.

• We pioneer the integration of DDPM with FMT, enabling high-quality FMT reconstruction. MDiff-FMT employs a novel noise-addition and denoising mechanism that enables precise morphological feature reconstruction through a multi-step probabilistic sampling process, significantly advancing the state-of-the-art in FMT reconstruction.

• We innovatively incorporate fluorescence signals as structural prior information into the MDiff-FMT framework, establishing an implicit mapping relationship between fluorescence signals and the underlying anatomical structures. This integration ensures a high degree of spatial consistency between the reconstructed images and the fluorescence signal distribution, significantly enhancing the reconstruction accuracy.

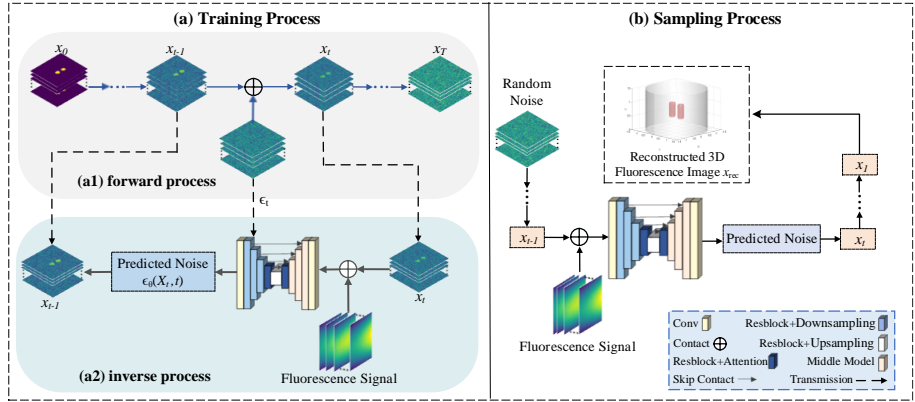

**Fig. 1.** Overview of the MDiff-FMT. (a)Training process includes (a1) forward process and (a2) inverse process. (b) Sampling process.

The rest of the paper is organized as follows. We introduce the proposed MDiff-FMT method in Section 2, present the experimental setup and results in Section 3, and conclude in Section 4.

## 2 Methodology

### 2.1 FMT Inverse Problem

FMT aims to reconstruct the 3D distribution of fluorophores $X$ from surface photon intensity measurements $\Phi$ using the diffusion equation (DE) [23]. The linear relationship between $\Phi$ and $X$ is modeled as:

$$\Phi = WX \tag{1}$$

where $W$ is the weight matrix derived from finite element analysis (FEA) [24]. Traditional methods [25] usually transform Eq. 1 into a minimization problem with a regularization term to obtain an approximate solution:

$$\min_x \{\frac{1}{2} \| Wx - \Phi \|_2^2 + \alpha R(x)\} \tag{2}$$

where $\alpha$ is the parameter that balances the data term and the regularization term, and $R(x)$ is the regularization term. Deep learning algorithms usually establish the relationship between $x$ and $\Phi$ by training an end-to-end mapping network [14]:

$$x_{net} = f_{net}(\Phi_{net}) \tag{3}$$

where $f_{net}$ is the constructed network model, $\Phi_{net}$ and $x_{net}$ are the input and output of the network, respectively. The model demonstrates excellent capability in establishing the mapping relationship between fluorescence signals and reconstructed images.

However, its performance heavily relies on thousands of training datasets to achieve excellent reconstruction quality.

### 2.2 MDiff-FMT for FMT Reconstruction

The network framework of the MDiff-FMT is shown in Fig. 1, which mainly includes the forward process $q(x_t | x_{t-1})$ and the inverse process $p_\theta(x_{t-1} | x_t)$. The forward process starts with the real label image $x_{ori}$ and gradually adds noise to obtain the Gaussian noise image $x_T$. In the inverse process, the reconstructed 3D image is progressively recovered by leveraging fluorescence signals as structural prior information. In the sampling stage, given the random noise and fluorescence signals, the well-trained model is used to generate a high-quality reconstructed image.

**Diffusion Forward Process.** In the forward process, Gaussian noise $\epsilon_t$ with variance $\beta_1,...\beta_T$ is gradually added to the real image $x_0 = x_{ori}$ to obtain a series of noise-added images $x_1,...x_T$. This process simulates the progressive degradation of morphological features caused by noise contamination, thereby providing a comprehensive spectrum of training samples that encompass varying degrees of structural detail preservation. The latent variable $x_t$ is defined as:

$$x_t = \sqrt{1-\beta_t}x_{t-1} + \sqrt{\beta_t}\epsilon_t, \ \epsilon_t \sim N(0,\mathbf{I}) \tag{4}$$

where $\mathbf{I}$ is an n*n dimensional unit matrix. The forward process can be viewed as a Markov chain, denoted as:

$$q(x_t | x_0) = \prod_{t=1}^{T} q(x_t | x_{t-1}) = \mathcal{N}(x_t; \sqrt{\bar{\alpha}_t}x_{ori}, (1-\bar{\alpha}_t)\mathbf{I}) \tag{5}$$

where $1 \leq t \leq T$, $x_T \sim \mathcal{N}(0,\mathbf{I})$ is a purely noisy image. $\alpha_t = 1-\beta_t$ and $\bar{\alpha}_t = \prod_{m=1}^{t} \alpha_m$. By reparameterization, Eq. 5 can be expressed as:

$$x_t = \sqrt{\bar{\alpha}_t}x_{ori} + \sqrt{(1-\bar{\alpha}_t)}\epsilon_t, \ \epsilon_t \sim N(0,\mathbf{I}) \tag{6}$$

Therefore, in the forward process, we initialize the procedure with the real image $x_{ori}$ and generate a series of noisy images $\{x_{ori}, x_1, x_2,... x_T\}$ and random noises $\{\varepsilon_0, \varepsilon_2, \varepsilon_3,...\varepsilon_T\}$ through Eq. 6. This carefully designed process establishes the essential training data foundation for the subsequent inverse reconstruction phase.

**Diffusion Inverse Processes.** The inverse process $p_\theta(x_{t-1} | x_t)$ starts with a noise and gradually removes the noise to obtain an image close to the original distribution [22]:

$$p_\theta(x_{t-1} | x_t) = \mathcal{N}(x_{t-1}; \mu_\theta(x_t,t), \tilde{\beta}_t\mathbf{I}) \text{ and } \tilde{\beta}_t = \frac{1-\bar{\alpha}_{t-1}}{1-\bar{\alpha}_t}\beta_t \tag{7}$$

$$\tilde{\mu}_\theta(x_t,t) = \frac{1}{\sqrt{\alpha_t}}(x_t - \frac{\beta_t}{\sqrt{1-\bar{\alpha}_t}}\epsilon_\theta(x_t,t)) \tag{8}$$

where $\epsilon_\theta(x_t,t)$ is the noise value at time t predicted by the neural network. After calculating mean value $\mu_\theta(x_t,t)$ by Eq. 8, we can predict $x_{t-1}$ by the following equation:

$$x_{t-1} = \mu_\theta(x_t,t) + \sqrt{\tilde{\beta}_t}z \quad z \sim N(0,\mathbf{I}) \tag{9}$$

Due to the uncertainty of $z$ and $\epsilon_\theta(x_t,t)$, the reconstructed image obtained according to Eq. 9 is random and has some error compared to the real label image $x_{ori}$. To obtain a reconstructed image that is consistent with $x_{ori}$, we limit the noise to be removed at each step by using the fluorescence signals $\Phi_c$ as structural information in the training process. The specific steps are as follows:

First, we combine the fluorescence signals $\Phi_c$ (size $(num,n,n)$) and the latent variable $x_t$ (size $(b,n,n)$) to obtain image $X_t$ of size $(num+b,n,n)$:

$$X_t = \Phi_c \oplus x_t \tag{10}$$

where $num$ is the number of fluorescence signals, and $b$ is the number of slices of $x_{ori}$. For $X_t$, we only update $x_t$ according to Eq. 6, while $\Phi_c$ remains unchanged.

Then, we input $X_t$ into the noise estimation model $\epsilon_\theta$, which employs a U-shaped network with residual blocks and temporal embedding modules [22]. The loss function of the model is set as:

$$E_{X_t,\epsilon_t,t}[\|\epsilon_t - \epsilon_\theta(X_t,t)\|^2] \tag{11}$$

where $\epsilon_t$ comes from the noisy data constructed by the forward process and $\epsilon_\theta(X_t,t)$ is the noise information to be removed at each step. After obtaining $\epsilon_\theta(X_t,t)$, the posterior distribution that $x_{t-1}$ obeys when $x_t$ is known is:

$$x_{t-1} = \mu_\theta(X_t,t) + \sqrt{\tilde{\beta}_t}z = \frac{1}{\sqrt{\alpha_t}}(x_t - \frac{\beta_t}{\sqrt{1-\bar{\alpha}_t}}\epsilon_\theta(X_t,t)) + \sqrt{\tilde{\beta}_t}z \quad z \sim N(0,\mathbf{I}) \tag{12}$$

**Sampling Process.** For the sampling process, the fluorescence signals $\Phi_c$ is incorporated as structural prior information to generate the FMT reconstructed image from the random noise image. Specifically, the initial noise map $X_t$ is first computed using Eq. 10, and then the reconstructed 3D fluorescence image $x_{rec}$ is obtained by iterating $T$ - steps through Eq. 12.

## 3      Experiments and Results

To fully evaluate the effectiveness of the proposed algorithm, we conducted experiments using a small-scale dataset of only 1,000 training samples through computer simulation. The reconstruction quality was assessed through Dice [26], contrast-to-noise ratio (CNR) [27], and localization error (LE) [12]. For fair comparison, MDiff-FMT, 3D-CNN and TransUnet were trained using the same dataset. MDiff-FMT, 3D-CNN and TransUnet were optimized using Adam with learning rate set to 1e$^{-4}$, batch size to 32, and epochs set to 10000, 3000, and 800, respectively. The number of iterations was set to 10,000 and 20 for the ART and StOMP, respectively, and the regularization parameter was set to 0.001 for the ART and the threshold was set to 0.8 for the StOMP.

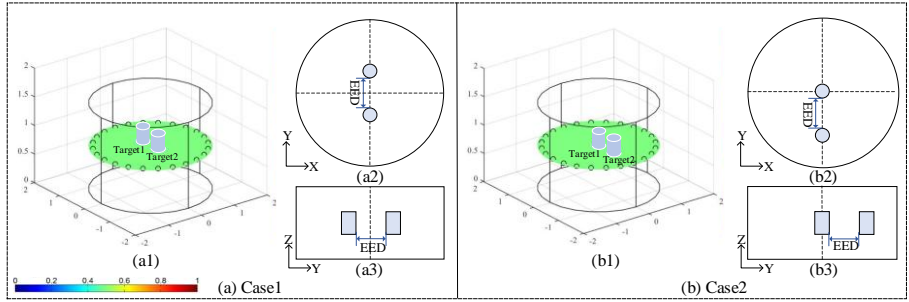

**Fig. 2.** Experimental setup in the numerical simulation for asymmetric case. (a) Case1: Symmetric dual-target configuration. (b) Case 2: Asymmetric dual-target configuration.

### 3.1 Numerical Simulation Experiments and Results

**Numerical Experimental Setting.** A simulated cylinder with a radius of 1.5 cm and a height of 1.5 cm was constructed, within which two fluorescent targets, each with a radius of 0.15 cm and a height of 0.5 cm, were placed at different locations. The XY planes, side planes, and 3D displays of fluorescent targets at two different positions are shown in Fig. 2, with EED values of 1mm, 2mm, and 3mm.

**Comparison with the State-of-the-art Methods.** To verify the superiority of the MDiff-FMT, we conducted comparative experiments with the classic iteration-based ART, StOMP, learning-based cutting-edge 3D-CNN, and TransUnet models. The experimental results are shown in Fig. 3. Obviously, ART cannot distinguish two fluorescent targets at all EEDs in both cases. StOMP and TransUnet can distinguish fluorescent targets at all EEDs, but the reconstructed targets exhibit irregular shapes. In Case 1, the shape of the fluorescent targets reconstructed by the 3D-CNN display irregular edge morphology. In Case 2, at EEDs of 1 mm and 2 mm, the two fluorescent targets reconstructed by the 3D-CNN display significant misalignment and cannot be distinguished from one another. In contrast, the MDiff-FMT successfully reconstructed the two fluorescent targets with more accurate positions and regular shapes in both cases. This result demonstrates that the MDiff-FMT achieves the best reconstruction performance in morphological restoration.

Furthermore, we quantitatively evaluated the results of the reconstruction of the five methods. As shown in Table 1, MDiff-FMT obtains an overall smallest lower LE, indicating that the proposed method reconstructs the positions of the fluorophores more accurately than the other four methods. In addition, compared with other four methods,

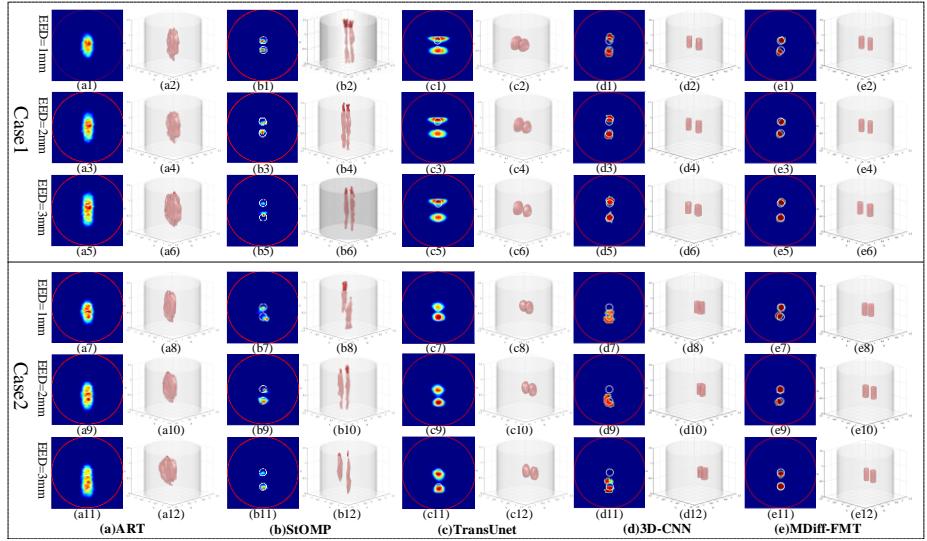

**Fig. 3.** Numerical simulation reconstruction results of five methods for different EEDs (i.e., 1 mm, 2 mm, 3 mm) in two cases. (a) ART ((a1) - (a12)), (b) StOMP ((b1) - (b12)), (c) TransUnet (c1) - (c12)), (d) 3D-CNN ((d1) - (d12)), (e) MDiff-FMT ((e1) - (e12)).

MDiff-FMT achieves the highest CNR and Dice at all EEDs in both cases. This indicates that the proposed MDiff-FMT achieves better signal-to-noise reconstruction and outperforms the other four methods in morphological recovery of fluorophores.

**Table 1.** Quantitative metrics for numerical simulation experiments

| Cases | Method | EED (1mm/2mm/3mm) | | | |
|---|---|---|---|---|---|
| | | CNR | LE1 (cm) | LE2 (cm) | Dice |
| Case1 | ART | 8.65/7.85/7.14 | 0.03/0.02/0.02 | 0.02/0.02/0.02 | 0.46/0.46/0.47 |
| | StOMP | 3.69/5.04/3.41 | 0.01/0.02/**0.01** | **0.01**/**0.01**/0.02 | 0.46/0.46/0.47 |
| | 3D-CNN | 6.78/8.34/12.07 | 0.02/0.03/0.03 | 0.02/0.02/0.02 | 0.41/0.50/0.61 |
| | TransUnet | 9.13/10.34/10.95 | 0.01/0.01/**0.01** | **0.01**/**0.01**/0.01 | 0.42/0.48/0.50 |
| | **MDiff-FMT** | **9.57/12.07/13.34** | **0.01/0.01**/0.02 | 0.02/0.02/**0.01** | **0.50/0.60/0.64** |
| Case2 | ART | 8.76/7.91/7.42 | 0.03/0.03/0.03 | 0.02/0.02/0.01 | 0.47/0.46/0.46 |
| | StOMP | 4.98/7.21/6.99 | **0.02**/0.02/**0.01** | 0.02/0.02/0.01 | 0.47/0.46/0.46 |
| | 3D-CNN | 5.88/4.53/4.42 | 0.03/0.03/0.03 | 0.02/0.02/0.02 | 0.16/0.18/0.29 |
| | TransUnet | 11.16/14.12/13.00 | 0.03/0.01/**0.01** | **0.01**/0.01/0.01 | 0.48/0.69/0.61 |
| | **MDiff-FMT** | **12.10/15.24/13.52** | 0.03/**0.01**/0.02 | 0.02/**0.01/0.01** | **0.50/0.71/0.64** |

### 3.2 Physical Phantom Experiments and Results

To further evaluate the performance of the Mdiff-FMT, we performed physical mimicry experiments in a custom-made FMT imaging system [14]. A cylindrical phantom, filled with a mixture of pure water and 1% intralipid, was used as the physical phantom (scattering and absorption coefficients of 10 cm$^{-1}$ and 0.02 cm$^{-1}$), with a diameter of 6 cm and a height of 3 cm. Two glass-like tubes (0.15 cm in radius) containing 20 μL of indocyanine green (ICG) at a concentration of 1.3 μM were immersed in the phantom as fluorescent targets, with an EED of 1 mm between them.

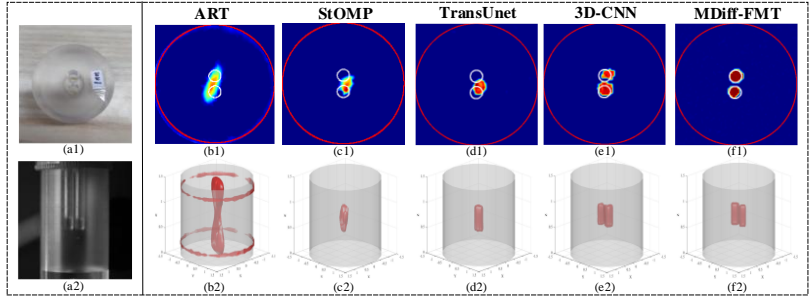

**Fig. 4.** The phantom experimental reconstruction results. (a1) – (a2) The physical phantom configurations. The 2D cross-sections and 3D renderings of the ART ((b1) - (b2)), StOMP ((c1) - (c2)), TransUnet ((d1) - (d2)), 3D-CNN ((e1) - (e2)), MDiff-FMT ((f1) - (f2)).

The reconstruction results of the five methods are shown in Fig. 4 (b-f) and Table 2. As illustrated in Fig. 4, the ART and StOMP methods cannot distinguish two fluorescent targets. TransUnet, which depends on large-scale datasets, is unable to distinguish two fluorescence targets due to the limited size of the training dataset. The 3D-CNN can distinguish the two fluorescent targets; however, it produces irregularly shaped edges. In contrast, the MDiff-FMT obtains the two fluorophores with a more precise location and regular shape. Quantitative indicators also show (Table 2) that the proposed MDiff-FMT achieves the highest CNR and DICE compared with other reconstruction algorithms, indicating that the MDiff-FMT method achieves better performance in restoring the morphology of fluorescent targets.

**Table 2.** Quantitative indicators of the physical phantom experiment

| Method | CNR | LE1 (cm) | LE2 (cm) | Dice |
|---|---|---|---|---|
| ART | 1.77 | 0.03 | 0.03 | 0.48 |
| StOMP | 3.07 | 0.03 | 0.02 | 0.16 |
| 3D-CNN | 10.14 | 0.02 | 0.02 | 0.44 |
| TransUnet | 6.34 | 0.02 | 0.03 | 0.28 |
| **MDiff-FMT** | **17.58** | **0.01** | **0.02** | **0.64** |

## 4 Conclusion

We report for the first time a morphology-aware diffusion model for 3D FMT reconstruction. The MDiff-FMT model employs a stepwise probabilistic sampling process to achieve the fine-grained reconstruction of morphological features. Furthermore, it incorporates fluorescence signals as structural priors to ensure accurate spatial correspondence between the reconstructed images and the fluorescence signals. Experimental results of simulation and phantom demonstrate that MDiff-FMT is capable of reconstructing high-quality 3D fluorescence images without the need for large-scale training data.

# References

1. Willmann, J. K., Van Bruggen, N., Dinkelborg, L. M., and Gambhir, S. S.: Molecular imaging in drug development. Nature reviews Drug discovery **7**(7): 591-607 (2008)
2. Stuker, F., Ripoll, J., and Rudin, M.: Fluorescence molecular tomography: principles and potential for pharmaceutical research. Pharmaceutics, **3**(2): 229-274 (2011)
3. Li, C., Zhou, S. S., Chen, J., and Jiang, X. Q.: Fluorescence Imaging of Inflammation with Optical Probes. Chemical and Biomedical Imaging **1**(6): 495-508 (2023)
4. He, Z. X., Zhang, L., Xing, L. X., Sun, W. J., Gao, X. J., Zhang, Y. Q., and Gao, F.: IR780‐based diffuse fluorescence tomography for cancer detection. Journal of Biophotonics **17**(5): 1-12 (2024)
5. Ntziachristos, V., Tung, C. H., Bremer, C., Weissleder R.: Fluorescence molecular tomography resolves protease activity in vivo. Nature Medicine **8**: 757–761 (2002)
6. Ntziachristos, V., Schellenberger, E. A., Ripoll, J., Yessayan, D., Graves, E., Bogdanov Jr, A., Josephson L., Weissleder, R.: Visualization of antitumor treatment by means of fluorescence molecular tomography with an annexin V–Cy5.5 conjugate. Proceedings of the National Academy of Sciences **101**(33): 12294-12299 (2004)
7. Cheng, J. J., and Luo, J. W.: Tikhonov-regularization-based projecting sparsity pursuit method for fluorescence molecular tomography reconstruction. Chinese Optics Letters **18**(1): 69-74 (2020)
8. Chen, Y., Du, M. F., Zhang, J., Zhang, G., Su, L. Z., Li, K., Zhao, F. J. Yi, H. J., Wang, L., and Cao, X.: Generalized conditional gradient method with adaptive regularization parameters for fluorescence molecular tomography. Optics Express **31**(11): 18128-18146 (2023)
9. Xie, W. H., Deng, Y., Wang, K., Yang, X. Q., and Luo, Q. M.: Reweighted L1 regularization for restraining artifacts in FMT reconstruction images with limited measurements. Optics letters **39**(14): 4148-4151 (2014)
10. Yuan, Y. T., Yi, H. J., Kang, D. Z., Yu, J. J., Guo, H. B., He, X. L., and He, X. W.: Robust transformed l1 metric for fluorescence molecular tomography. Computer methods and programs in biomedicine **234**: 107503 (2023)
11. Luo, X. L., Ren, Q. Q., Zhang, H., Chen, C., Yang, T., He, X. W., and Zhao, W.: Efficient FMT reconstruction based on L1–αL2 regularization via half-quadratic splitting and a two-probe separation light source strategy. Journal of the Optical Society of America A **40**(6): 1128-1141 (2023)
12. Guo, L., Liu, F., Cai, C. J., Liu, J., and Zhang, G. L.: 3D deep encoder–decoder network for fluorescence molecular tomography. Optic letters **44**(8): 1892-1895 (2019)
13. Meng, H., Gao, Y., Yang, X., Wang, K., and Tian, J.: K-nearest neighbor based locally connected network for fast morphological reconstruction in fluorescence molecular tomography. IEEE transactions on medical imaging **39**(10): 3019-3028 (2020)
14. Zhang, P., Fan, G. D., Xing, T. T., Song, F., Zhang, G. L.: UHR-DeepFMT: Ultra-High Spatial Resolution Reconstruction of Fluorescence Molecular Tomography Based on 3-D Fusion Dual-Sampling Deep Neural Network. IEEE Trans Med Imaging **40**(11): 3217-3228 (2021)
15. Cao, C. G., Xiao, A., Cai, M. S., Shen, B. L., Guo, L. S., Shi, X. J., J, T., and Hu, Z. H.: Excitation-based fully connected network for precise NIR-II fluorescence molecular tomography. Biomedical Optics Express **13**(12): 6284-6299 (2022)
16. Zhang, X. X., Jia, Y. F., Cui, J. P., Zhang, J., Cao, X., Zhang, L., and Zhang, G. L.: Two-stage deep learning method for sparse-view fluorescence molecular tomography reconstruction. Journal of the Optical Society of America A **40**(7): 1359-1371 (2023)

17. Hua, Y. Z., Jiang, Y. X., Liu, K. X., Luo, Q. M., and Deng, Y.: Interpretable model-driven projected gradient descent network for high-quality fDOT reconstruction. Optics Letters **47**(10): 2538-2541 (2022)
18. Sun, W. J., Zhang, L. M., Xing, L. X., He, Z. X., Zhang, Y. Q., and Gao, F.: Projected algebraic reconstruction technique-network for high-fidelity diffuse fluorescence tomography re-construction. Journal of the Optical Society of America A **41**(6): 988-999 (2024)
19. Yang, Y., Wan, W. B., and Zhou, H. L.: A model-driven deep unfolding network for fluorescence molecular tomography reconstruction. In: Optics in Health Care and Biomedical Optics XII, pp. 294-305. SPIE, online (2022)
20. Jiang, Y. X., Liu, K. X., Li, W. S., Luo, Q. M., & Deng, Y.: Deep background-mismodeling-learned reconstruction for high-accuracy fluorescence diffuse optical tomography. Optics Letters **48**(13): 3359-3362 (2023)
21. Liu, K., Jiang, Y., Li, W., Chen, H., Luo, Q., & Deng, Y.: High-fidelity mesoscopic fluorescence molecular tomography based on SSB-Net. Optics Letters **48**(2): 199-202 (2023)
22. Ho, J., Jain, A., and Abbeel, P.: Denoising Diffusion Probabilistic Models. arXiv:2006.11239 (2020)
23. Han, D., Tian, J., Zhu, S. P., Feng, J. C., Qin, C. H., Zhang, B., and Yang, X.: A fast reconstruction algorithm for fluorescence molecular tomography with sparsity regularization. Optics express **18**(8): 8630-8646 (2010)
24. Soubret, A., Ripoll, J., and Ntziachristos, V.: Accuracy of fluorescent tomography in the presence of heterogeneities: study of the normalized Born ratio. IEEE transactions on medical imaging **24**(10): 1377-1386 (2005)
25. Guo, H. B., Yu, J. J., He, X. W., Hou, Y. Q., Dong, F., and Zhang, S. L.: Improved sparse reconstruction for fluorescence molecular tomography with L 1/2 regularization. Biomedical optics express **6**(5):1648-1664 (2015)
26. Chen, C., Tian, F. H., Liu, H. L., and Huang, J. Z.: Diffuse optical tomography enhanced by clustered sparsity for functional brain imaging. IEEE transactions on medical imaging **33**(12): 2323-2331 (2014)
27. Zhang, G. L., Pu, H. S., He, W., Liu, F., Luo, J. W., and Bai, J.: Full-direct method for imaging pharmacokinetic parameters in dynamic fluorescence molecular tomography. Applied Physics Letters **106**(8). 2015.